# Sex Differences in 6-Year Progression of White Matter Hyperintensities in Non-Demented Older Adults: Sydney Memory and Ageing Study

Abdullah Alqarni [1,2,*], Wei Wen [1,3], Ben C.P. Lam [1], Nicole Kochan [1], Henry Brodaty [1], Perminder S. Sachdev [1,3], Jiyang Jiang [1,*]

[1] Centre for Healthy Brain Ageing (CHeBA), Discipline of Psychiatry and Mental Health, School of Clinical Medicine, Faculty of Medicine and Health, University of New South Wales, Sydney NSW, Australia

[2] Radiology and Medical Imaging Department, College of Applied Medical Sciences, Prince Sattam bin Abdulaziz University, Al-Kharj, Saudi Arabia.

[3] Neuropsychiatric Institute, Prince of Wales Hospital, Randwick, NSW, Australia.

*Corresponding Authors: Abdullah Alqarni (a.alqarni@student.unsw.edu.au); Jiyang Jiang (jiyang.jiang@unsw.edu.au); Centre for Healthy Brain Ageing (CHeBA), Level 1, AGSM building (G27), UNSW Gate 11, Botany Street, Kensington NSW 2052 Australia.

# Abstract

**Objectives:** To examine sex differences in the associations between vascular risk factors and 6-year changes in the volume of white matter hyperintensities (WMH), and between changes in WMH volumes and changes in cognitive performance, in a cohort of non-demented older adults.

**Methods:** WMH volumes at 3 time-points (baseline, and 2- and 6-year follow-up) were automatically quantified in participants of Sydney Memory and Ageing Study (N = 605; age range = 70-92 years; 54.78% female). Linear mixed models were applied to examine the effects of vascular risk factors and cognitive consequences of the progression of WMH, as well as the sex moderation effects in the associations.

**Results:** Total (TWMH), periventricular (PVWMH), and deep (DWMH) WMH volumes increased by 9.47%, 7.70%, and 11.78% per year, respectively. No sex differences were found in WMH progression rates. After Bonferroni correction, increases in PVWMH volumes over time were associated with decline in global cognition, especially in visuospatial and memory domains. Men with more increases in PVWMH volumes over time had greater declines in visuospatial abilities. Moreover, higher average TWMH volumes across time-points were associated with poorer average performance in processing speed and executive function domains across time. Higher average PVWMH volumes across time-points were also associated with worse average performance in the executive function domain over time, among women but not men.

**Conclusion:** The findings highlighted sex differences in the associations between WMH progression and cognition decline over time, suggesting sex-specific strategies in managing WMH accumulations in ageing.

# Introduction

White matter hyperintensities (WMH) are abnormally bright clusters on T2-weighted Magnetic Resonance Imaging (MRI) scans, which are presumed to be originated from vascular disorders, and considered as a biomarker for cerebral small vessel diseases (SVD) (Wardlaw et al., 2013). WMH are common in advanced ageing populations, and the severity of WMH is significantly higher in individuals with a diagnosis of dementia (Garnier-Crussard et al., 2020). Multiple vascular risk factors may contribute to the accumulation of WMH, including hypertension (Habes et al., 2018; Nyquist et al., 2015), obesity and blood lipid abnormalities (Alqarni et al., 2021; Griffanti et al., 2018; Lampe et al., 2019; Yin et al., 2018), diabetes (Raffield et al., 2016), smoking (Habes et al., 2018; Sachdev et al., 2009), and genetics (Armstrong et al., 2020; Persyn et al., 2020). WMH in the periventricular (PVWMH) and deep white matter (DWMH) regions were found to be affected differently by risk factors (Armstrong et al., 2020; Schmidt et al., 2011; Wharton et al., 2015), warranting separate analyses for these two regions (Griffanti et al., 2018).

Women were found to have greater WMH volumes and faster accumulations of WMH over time, compared to men (de Leeuw et al., 2001; DeCarli et al., 2005; Fatemi et al., 2018; Sachdev et al., 2009; van den Heuvel et al., 2004). Nevertheless, the associations between WMH volumes and some vascular risk factors, including hypertension (Assareh et al., 2014; Filomena et al., 2015) and atherosclerosis (Geerlings et al., 2010), were reported in men. Moreover, although obesity was associated with increased WMH volumes in both men and women (Griffanti et al., 2018; Lampe et al., 2019; Veldsman et al., 2020), the effects were greater in men (Alqarni et al., 2021). On the other hand, the associations between diabetes and WMH were only found in women but not in men (Espeland et al., 2019; Jongen et al., 2007). Collectively, these findings suggest a significant sex dimorphism in the effects of vascular risk factors on WMH volumes. Men were found to be more likely to develop mild cognition impairment (MCI) with faster WMH progression rate (Burke et al., 2019), although some studies showed that higher WMH were associated with higher risks of amnestic MCI in women, but not in men (Debette et al., 2010). More WMH in the frontal lobe were associated with higher intra-individual reaction time variability in women, whereas more WMH in the temporal lobe were associated with declined face recognition function in men (Bunce et al., 2010).

WMH start to appear on brain MRI scans in middle-aged individuals in their forties (Wen et al., 2009), and can be found in the majority (90%) of community-based individuals by the age of 80 (de Leeuw et al., 2001). In a longitudinal study with older adults aged 62.5±7.7 years at baseline, WMH were found to increase 4.7 mL in 9 years (van Leijsen et al., 2017). Longitudinal studies have associated greater WMH accumulations with older age, diagnosis of dementia, baseline hypertension, and higher cardiovascular disease risk scores (Carmichael et al., 2010)(Scharf et al., 2019). The progression of WMH in women was associated with baseline hypertension, whereas the increase of WMH volumes in men was related to higher baseline fasting glucose, baseline hypertension, and midlife hypertension (Scharf et al., 2019). Meanwhile, some studies showed a decrease of WMH volumes over time. van Leijsen *et al.* (van Leijsen et al., 2017) found that 9.4% of the whole sample had reduced WMH volumes in the 1-year follow-up. Another study found decreased WMH volumes in 37% of patients 1 year after a minor stroke, which was accompanied by significantly reduced blood pressure (Wardlaw et al., 2017).

To our knowledge, sex differences in the associations between vascular risk factors and WMH, and between WMH and cognition, have not been comprehensively investigated in a longitudinal study design. Therefore, in this study, we aim to examine these sex dimorphisms in a longitudinal study of 605 older adults (3 time-points over 6 years). Examining sex differences will provide insights for better understanding of different WMH accumulation mechanisms in men and women and deliver evidence to plan for sex-specific management of WMH progression in clinical settings.

## Methods

### Study sample

The study sample was drawn from baseline, and 2- and 6-year follow-ups of Sydney Memory and Ageing Study (MAS), a community-based longitudinal study of older adults (baseline age range, 70-90 years; (Sachdev et al., 2010)). Individuals with a diagnosis or current evidence of dementia, mental retardation, psychotic disorder including schizophrenia or bipolar disorder, multiple sclerosis, motor neuron disease, developmental disability, or progressive malignancy, were excluded at baseline. One thousand and thirty-seven participants were finally included in the baseline of MAS. Six hundred and five participants with MRI data for WMH quantification from at least one time-point were included in the current study (Table 1).

### MRI acquisition

Baseline MRI scans were acquired on two MRI scanners, a Philips 3T Intera Quasar scanner and a Philips 3T Achieva Quasar Dual scanner (Philips Medical Systems, The Netherlands). Identical MRI scanning parameters were used on these two scanners to acquire T1-weighted and T2-weighted Fluid Attenuated Inversion Recovery (FLAIR) scans. Specifically, for T1-weighted imaging, the following parameters were used: repetition time (TR) = 6.39 ms, echo time (TE) = 2.9 ms, flip angle = 8°, matrix size = 256×256, field of view (FOV) = 256×256×190, and slice thickness = 1 mm with no gap in between, yielding 1×1×1 mm$^3$ isotropic voxels. Imaging parameters for the T2-weighted FLAIR sequence include TR = 10,000 ms, TE = 110 ms, inversion time (TI) = 2800 ms, matrix size = 512×512, slice thickness = 3.5 mm without gap, and in-plane resolution = 0.488×0.488 mm. All MRI scans at the 2- and 6-year follow-up were acquired from the Philips 3T Achieva Quasar Dual scanner with the same imaging parameters. In our statistical analyses, a dummy variable was included to adjust for the potential effects introduced by different scanners.

Vascular risk factors

Vascular risk factors examined in this study include the diagnosis of hypertension (HT) and diabetes by clinicians, smoking (ever smoked tobacco regularly (Yes/No)), alcohol consumption (≤ 4 intakes per month = low consumption; ≥ 2 intakes per week = high consumption), hip to waist ratio (HWR), systolic (SBP) and diastolic (DBP) blood pressure, body mass index (BMI), total cholesterol (TC), triglycerides, high density lipoprotein (HDL), low density lipoprotein (LDL), and pulse rate (Sachdev et al., 2010). In addition to examining individual risk factors, the composite Framingham cardiovascular disease (CVD) risk score was also calculated through combining smoking, diabetes, SBP, HDL, TC, and BMI (D'Agostino et al., 2008). The purposes for the inclusion of Framingham CVD risk score in addition to individual risk factors are: 1) to investigate the effects of overall vascular risks on the progression of WMH, and 2) to put the current results in the context of previous clinical studies where the Framingham risk score was widely used.

### WMH segmentation

WMH volumes were automatically quantified by using UBO Detector (Jiang et al., 2018). The longitudinal module of UBO Detector was applied for data acquired at multiple time-points. In the longitudinal module, the T1-weighted scan acquired at the first time-point was used as the reference. All other T1 and FLAIR images of the same participant were linearly registered to the reference. Tissue segmentations and warping to Diffeomorphic Anatomical Registration Through Exponentiated Lie (DARTEL) (Ashburner, 2007) space were conducted on the reference T1 image. The resultant flow field map was then applied to all other T1 and FLAIR to transform them to the same DARTEL space. To classify candidate clusters generated by FSL FAST into WMH vs. non-WMH, a *k*-nearest neighbours classifier was applied by considering intensity, anatomical location, and cluster size features. PVWMH was defined as WMH voxels with a distance of < 12 mm from the boundary of lateral ventricles. Other WMH voxels were classified as DWMH. Standard UBO Detector processing steps were applied to data acquired at a single time-point. Visual quantity control was conducted by superimposing generated WMH masks onto corresponding FLAIR scans. Thirteen scans were removed due to poor image quality or inaccurate WMH segmentation results. Since WMH were segmented and quantified in a standard space (DARTEL space), differences in intracranial volumes between participants do not need to be adjusted for in statistical analyses.

### Neuropsychological assessments

Comprehensive neuropsychological assessment batteries were administered to MAS participants to examine their performance in five cognitive domains, namely processing speed, memory, language, visuospatial function, and executive function (Sachdev et al., 2010). Briefly, for the *processing speed* domain, the Wechsler Adult Intelligence Scale-III Digit Symbol-Coding (Wechsler, 1997a) and Trail Making Test part A (Strauss et al., 2006) were used. To evaluate *memory* function, Logical Memory Story A (delayed recall) (Wechsler, 1997b), Rey Auditory Verbal Learning Test (Strauss et al., 2006), and Benton Visual Retention Test recognition (Benton AL, 1996) tests were used. Boston Naming Test (30 items) (Kaplan, 2001) and Semantic Fluency Test (animals) (Strauss et al., 2006) were applied to assess performance in the *language* domain. Block Design from the Wechsler Adult Intelligence Scale—Revised (Wechsler, 1981) was used to assess *visuospatial* function. Controlled Oral Word Association Test and Trail Making Test part B (Strauss et al., 2006) was used to assess the *executive function* domain. Z-scores for each test were calculated for each cognitive domain, by using the means and standard deviations derived from a healthy subsample at baseline. Domain z-scores were calculated by averaging the component tests and standardising against the healthy subsample. Finally, to calculate the *global cognition* scores, domain z-scores were averaged and standardized.

Statistical analysis

IBM SPSS 27.0 (IBM Corp. Released 2020. IBM SPSS Statistics for Windows, Version 27.0. Armonk, NY: IBM Corp) was used to conduct the statistical analyses. WMH volumes (when used as dependent variables in the model) were transformed using the natural logarithm due to the non-Gaussian distribution of the raw WMH volumes. Linear mixed models were then fitted to examine 1) changes in WMH volumes over time, 2) effects of risk factors on changes in WMH volumes, 3) associations of WMH volume changes and changes in cognitive performance, and 4) sex moderation effects on these associations.

*First*, the changes in WMH volumes over time were examined by including transformed WMH as the outcome, and time-in-study (TIS) in years, baseline age, scanner, and sex as predictors. Random effects of the intercept and TIS were also included in the model. The TIS term assesses annual changes in transformed WMH volumes. Sex differences in annual WMH change rates were examined by further including the interaction term between TIS and sex (TIS * sex) into the model. A significant TIS * sex interaction indicates that there are significant sex differences in annual WMH changes.

*Second*, the main effects of vascular risk factors at baseline on WMH volume changes over time were examined by including the interaction between each individual baseline risk factor and TIS (TIS * risk factor), baseline risk factor, TIS, baseline age, scanner, and sex, into the model as predictors, and the transformed WMH volume as the outcome. Each individual risk factor and CVD risk score were examined in separate models. A significant TIS * risk factor interaction indicates that the baseline risk factor is associated with the annual change rate of transformed WMH volumes. Sex differences in the effects of baseline vascular risk factors on WMH changes were examined by further including an interaction term between TIS, sex and risk factor (TIS * risk factor * sex) into the model. A significant TIS * risk factor * sex interaction indicates that the association between the risk factor and annual changes in transformed WMH volumes differs between men and women. For significant TIS * risk factor * sex interactions, we then examined the TIS * risk factor term in men and women separately.

*Third*, the associations between changes in WMH volumes and changes in cognitive performance were examined by using the method described in (Hedeker and Gibbons, 2006).

Specifically, the model incorporated 1) an individual-mean centred, time-varying WMH variable, capturing within-subject variation in WMH volumes over time, and 2) an individual's mean WMH volumes across time-points, capturing between-subject differences in WMH volumes. In addition, TIS, baseline age, scanner, sex, and baseline years of education, were included in the model as predictors. Scores from neuropsychological tests (i.e., global cognition or domain cognitive scores) were used as the outcome variables. Sex differences in the between- and within-subject effects of WMH progression on changes in cognition were examined by further including interaction terms (sex * between-subject WMH differences, sex * within-subject WMH differences) in the models.

In all models, continuous risk factors were grand-mean centred by subtracting the overall means, except for the within-subject WMH difference described above. In addition, variables with relatively small units were rescaled or converted to larger units to better illustrate their effects. Specifically, SBP and DBP values were divided by 10 and interpreted with units of 10 mmHg. Moreover, WMH volumes quantified by UBO Detector were converted from $mm^3$ to mL by dividing the WMH volumes in $mm^3$ by 1000 (1 $mm^3$ = 0.001 mL) to better describe their effects on cognition. To adjust for multiple testing, Bonferroni correction was applied to the models investigating the associations between risk factors and WMH (the first and second sets of models), and the associations between WMH and cognition (the third set of models). The first and second sets of models consisted of 3 outcomes (TWMH, PVWMH, DWMH) and 2 types of analyses (main and interaction effects). Therefore, *p*-values less than 0.05/6 = 0.00833 were considered surviving Bonferroni correction. The third set of models consisted of 3 predictors (TWMH, PVWMH, DWMH) and 6 cognition variables. Therefore, *p*-values less than 0.05/18 = 0.00277 were considered significant. For the unstandardised B generated from the linear mixed models with transformed WMH as the outcome variable, exponentiated B (i.e., exp(B)) was calculated to report the percentage change in the outcome variable associated with a unit change in the predictor. When exp(B) > 1, results were interpreted as a unit increase in the predictor being associated with ((exp(B)-1)*100 percent increase in the outcome variable. When exp(B) < 1, results were interpreted as a unit increase in the predictor being associated with a (1 - exp(B)) * 100 percent decrease in the outcome variable.

## Results

### Sample characteristics

Table 1 summarises sample characteristics. At baseline, women had higher TWMH and DWMH volumes, compared to men. Men had more years of education, and higher BMI, HWR, and SBP, whereas women had higher TC, HDL, LDL, and pulse rate. There were higher proportion of men with diabetes ($p = 0.011$; men, 12.88%; women, 8.42%; whole sample, 10.42%). There were also higher proportion of men who were regular smokers ($p < 0.001$; men, 63.14%; women, 42.81%; whole sample, 52.02%), and consumed ≥ 2 intakes of alcohol per week ($p < 0.001$; men, 68.64%; women, 45.45%; whole sample, 55.94%). Differences in risk factors and cognitive test scores between men and women at the 2- and 6-year follow-up are summarised in Supplementary Table 1.

### The progression of WMH over time

After adjusting for baseline age, sex, and scanner, TWMH volumes showed an increase rate of 9.47% per year ($p < 0.001$), with relatively lower increase rate for PVWMH (7.70% / year; p

< 0.001) than DWMH (11.78% / year; p < 0.001) (Figure 1). Men and women did not show significant differences in WMH progression rates (TWMH, p = 0.385; PVWMH, p = 0.545; DWMH, p = 0.098).

### The effects of vascular risk factors on the progression of WMH

The effects of vascular risk factors on changes in WMH volumes over time are summarised in Table 2. Participants with one-unit higher HWR at baseline showed 10.5% less annual increase of PVWMH volumes, but 22.3% more annual increase of DWMH volumes. One mmol/L higher HDL at baseline was associated with 1.8% more annual increase of PVWMH volumes. An additional 10 mmHg of SBP at baseline was associated with 0.5% less annual increase of PVWMH and 1% less annual increase of DWMH volumes. Participants with 10 mmHg higher DBP at baseline had 1.7% less annual increase of DWMH volumes. The effects of other risk factors at baseline (BMI, TC, Triglycerides, LDL, HT, pulse rate, diabetes, smoking, alcohol, and CVD risk score) on WMH volume changes over time were not statistically significant. None of the associations survived Bonferroni correction.

### Sex differences in the effects of vascular factors on the progression of WMH

The moderation effects of sex on associations between risk factors and WMH volume changes are summarised in Table 3. The interaction between TIS, sex, and baseline SBP was statistically significant in the models where TWMH (p = 0.040) and DWMH (p = 0.020) volume changes were the outcome variables. Ten mmHg higher baseline SBP were associated with 1.2% less annual increase of TWMH and 2.1% less annual DWMH accumulations in men. There were no significant effects of SBP on WMH changes in women. The sex moderation effects in associations between SBP and WMH did not survive Bonferroni correction.

Moreover, the interaction between TIS, sex, and baseline CVD risk scores was significant in the model where the DWMH volume change was the outcome variable ($p = 0.006$). One point higher Framingham CVD risk score at baseline was linked to 1.3% less annual increase of DWMH volumes in men. The effects of Framingham CVD risk scores on WMH changes over time were not significant in women. The moderation effects of sex on the association between CVD risk score and DWMH changes survived Bonferroni correction. In order to investigate the unexpected association between higher baseline CVD risk scores and less increase of DWMH burdens, the study sample was separated into young and old subsamples using the median age of 79.107 years. Significant sex moderation effects on the association between CVD risk scores and DWMH changes were only found in the older subsample ($p = 0.016$; Supplementary Table 2); older men with one-unit higher CVD risk score had 1.9% less DWMH increase each year ($p = 0.003$). Independent sample *t*-tests showed significantly higher CVD risk scores in old (mean ± standard deviation (SD), $4.418 \pm 3.151$; 45.60% male), compared to young (mean ± SD, $3.932 \pm 3.053$; 44.59% male), participants in the whole sample ($t = -3.297$, $p = 0.001$). However, no significant differences in CVD risk scores were found between young and old men ($t = -0.268$, $p = 0.789$).

The associations between WMH progression and changes in cognitive performance

The associations between changes in WMH and changes in cognitive performance are summarised in Table 4. More increase of PVWMH volumes over time (i.e., within-subject effect) was associated with more decline in executive function ($B = -0.022$, $p = 0.040$), visuospatial ability (B = -0.029, p = 0.001), and memory (B = -0.027, p = 0.003) domains, as well as global cognition (B = -0.026, p = 0.001) across time. More increase in TWMH volumes (within-subject effect) was also associated with more decline in memory across time (B = -0.005, p = 0.029). The associations of PVWMH with visuospatial abilities, memory, and global cognition (within-subject effects) remained significant after Bonferroni correction.

Moreover, higher average TWMH, PVWMH and DWMH volumes across time-points (i.e., between-subject effects) were associated with poorer average performance in processing speed (B = -0.014 ~ -0.006, p ≤ 0.005) and executive function (B = -0.016 ~ -0.007, p ≤ 0.004), and worse average global cognition (B = -0.014 ~ -0.005, p ≤ 0.033) across time-points. After Bonferroni correction for multiple testing, the associations of TWMH with processing speed and executive function (between-subject effects), the association between PVWMH and executive function (between-subject effect), remained statistically significant.

Sex differences in the effects of WMH progression on changes in cognitive performance

Sex differences in the between-subject and within-subject effects of WMH changes on changes in cognition are summarised in Table 5. Sex moderation effects on the associations between within-subject changes in TWMH (p = 0.022) and PVWMH (p = 0.002) volumes (i.e., within-subject effect) and changes in visuospatial ability were also significant, where men, but not women, with more increase in TWMH and PVWMH volumes had more decline in visuospatial function. In addition, sex moderation effects on the relationship between average PVWMH volumes across time points (i.e., between-subject effect) and average performance in executive function, were statistically significant (p= 0.002), where women, but not men, with higher average PVWMH volumes across time-points showed poorer average performance in executive function ($p < 0.001$). After Bonferroni correction, the sex moderation effects on the associations between PVWMH and visuospatial abilities (within-subject effects), and between PVWMH and executive function (between-subject effects), remained significant.

## Discussion

Sex differences in the impacts of vascular risk factors and cognitive consequences of 6-year WMH progression were examined in a cohort of 605 non-demented older adults aged 70-92 years at baseline. Results showed that TWMH increased at an annual rate of 9.47%, with a higher annual increase rate in DWMH (11.78%) than that in PVWMH (7.70%). After Bonferroni correction, none of the individual vascular risk factors examined significantly contributed to the progression of WMH. The association between Framingham CVD risk scores and DWMH significantly differed between men and women, where men, but not women, with higher CVD risk scores had less DWMH accumulations over time. The increase of PVWMH volumes over time was associated with declines in visuospatial and memory domains, as well as global cognition (within-subject effects). Higher average TWMH volumes across time-points were associated with worse average performance across time-points in both processing speed and executive function (between-subject effects). Higher average PVWMH volumes across time-points were also associated with worse average performance across time-points in executive function (between-subject effects). When sex differences in the cognitive consequences of WMH were examined, the within-subject effects of PVWMH increase on declined performance in visuospatial domain were only found in men, whereas the between-subject effects of average

PVWMH volumes on average performance in executive function domain were only found in women.

Previous studies have shown the increase of WMH volumes (Carmichael et al., 2010; Scharf et al., 2019; van Leijsen et al., 2017), and that women had higher increase rate of WMH compared to men (de Leeuw et al., 2001; van den Heuvel et al., 2004). Limited evidence on sex effects on the progression of WMH showed faster progression of DWMH in women than men ((van den Heuvel et al., 2004); N = 554; Age = 70-82). Although women were found to have higher TWMH and DWMH volumes at baseline, no significant sex differences in 6-year progression of WMH were found in this study. This inconsistency may be partially due to the inclusion of older participants in the current study. A previous study with an age range of 50–91 years found that hemodynamic impairment, such as reduced cerebrovascular reactivity, may accelerate the process of WMH accumulation over time (Sam et al., 2016). It was also found that the hormonal imbalance after menopause lowers neuroprotection and contribute to the development of WMH in women at middle age (Seo et al., 2013; Thurston et al., 2016; Thurston et al., 2022). These effects may collectively diminished sex differences.

Previous studies have shown the contributions of HT (Firbank et al., 2007; White et al., 2019), diabetes (de Bresser et al., 2010; de Havenon et al., 2019), the co-existence of other SVD biomarkers (Gouw et al., 2008; van Leijsen et al., 2017), and the presence of multiple vascular risk factors such as HT, BMI, diabetes, and smoking (Habes et al., 2016; van Dalen et al., 2017; Yoon et al., 2017), to WMH burdens. We also previously reported, in a cross-sectional study, that HT and HWR were independently associated with WMH, and the association between BMI and DWMH was moderated by sex (Alqarni et al., 2021). However, sex differences in the longitudinal associations between vascular risk factors and WMH have not been comprehensively documented in the literature. We found that men with higher CVD risk scores had less DWMH volume increase, and this relationship between CVD risk score and DWMH volume changes was not found in women. While this finding seems inconsistent with existing studies, we suspect this may be partially due to healthy survivor effects. Indeed, when stratifying the sample into young and old sub-groups using median age (79 years), the inverse relationship between CVD risk scores and DWMH volume changes was only significant in older men. Independent sample *t*-test showed no significant differences in CVD risk scores between young and old men, although the differences between young and older participants were significant in the whole sample. These findings suggest that older men included in this study tend to be healthy, and this may contribute to the result of sex differences in the relationship between CVD risk scores and DWMH volume changes. We should note that age and sex were not included in the computation of CVD risk score because they were included as covariates in all analyses.

The longitudinal associations between increased WMH and cognition decline have been previously reported. Processing speed decline over time was associated with the progression of TWMH, especially anterior WMH progression, while the progression of posterior WMH was associated with visual-constructional skills decline over time (Marquine et al., 2010). However, most studies did not separate between-subject from within-subject effects, making the interpretation inconclusive. In this study, we showed that higher average TWMH and PVWMH volumes were associated with poorer averaged cognitive performance in processing speed and executive function across time-points (between-subject effects). Greater within-subject increases in PVWMH volumes were associated with declines in visuospatial function and memory, and global cognition over time. Previous studies showed WMH-related decline in visuospatial function, but no sex differences were reported (Au et al., 2006; Stavitsky et al., 2010). The current study showed that between-subject effects of PVWMH on executive function were specific to women, whereas the within-subject effects of PVWMH on visuospatial ability were only found in men. A recent study showed that WMH progression in men have led to faster development of mild cognition impairment (MCI), whereas women, who were older and scored higher in the clinical dementia rating (CDR) scale at baseline, were at higher risk of developing MCI (Burke et al., 2019). In addition, a clinical trial that investigated the longitudinal associations between WMH progression and cognition decline risk using Alzheimer Disease Assessment Scale-Cognitive Subscale (ADAS-Cog), Mini-Mental State Examination (MMSE), and CDR scale, found that higher WMH volume at baseline and WMH accumulations were associated with decreased MMSE and higher ADAS-Cog in the 1-year follow-up (Carmichael et al., 2010). Contradicting results were found in another study, where WMH progression was not related to dementias, including Alzheimer's disease, Parkinson's disease, and dementia with Lewy bodies, or changes in global cognition over time (Burton et al., 2006). However, this study (Burton et al., 2006) had a much smaller sample size compared to other studies that found associations between the progression of WMH and cognitive decline over time (Carmichael et al., 2010).

In this study, we examined the time-change effects on regional WMH (PVWMH/DWMH). DWMH was found to have different aetiologies, genetic contributions, pathohistological traits, and extent of axonal loss, when comparing to PVWMH (Alqarni et al., 2021; Armstrong et al., 2020; Kim et al., 2008; Schmidt et al., 2011; Wardlaw et al., 2015; Wharton et al., 2015). Moreover, increased amyloid-β peptides deposition in leptomeningeal and cortical vessels walls was associated with DWMH (Gurol et al., 2020). Therefore, the research into WMH development would benefit from investigating PVWMH and DWMH separately and could provide better understanding of their progression mechanisms.

The current study has several strengths and limitations. WMH start to appear at middle age (Wardlaw et al., 2015). The current findings are more focused on a late age stage. At this late age stage, there might be a plateau of WMH progression caused by reduced effects of cardiovascular risk factors (Moscufo et al., 2012). Participants who were diagnosed with cardiovascular diseases or at high risk of developing one might have used medications to avoid the effects of cardiovascular diseases. However, no intended intensive treatment to reduce cardiovascular diseases risks was conducted for this study sample. Through separating within-subject from between-subject effects, the current study provides better understanding on the effects of WMH changes on cognitive decline over time. Future studies can link hormonal factors with WMH progression to examine the underlying mechanisms on the observed sex differences. Hormonal factors, such as menopause (Lohner et al., 2022) and hormone replacement therapy (HRT) (Liu et al., 2009), have been associated with WMH cross-sectionally. Our previous findings showed that menopause, testosterone, and HRT moderated the associations between vascular risk factors and WMH, in a cross-sectional study (Alqarni et al., 2022).

## Conclusion

This study examined sex differences in the progression of WMH in non-demented older individuals. Results highlighted that sex should be taken into consideration in the management of vascular risk factors for reducing WMH and improving performance in specific cognitive domains.

**Acknowledgement**

We would like to thank MAS participants and the research teams of the project. Memory and Ageing Study (MAS) was supported by the National Health and Medical Research Council (NHMRC) Program Grant ID 350833.

**Disclosure**

The authors report no disclosures relevant to the manuscript.

# Tables

## Table 1: Sample characteristics

| Parameter | Baseline | | | | 2-year follow-up | | 6-year follow-up | |
|---|---|---|---|---|---|---|---|---|
| | Female | Male | Statistics | p-value | Female | Male | Female | Male |
| N | 286 | 236 | - | - | 211 | 193 | 135 | 125 |
| Age (years) | 79.281 ± 4.680 | 79.182 ± 4.409 | 0.815[a] | 0.415 | 80.505 ± 4.680 | 80.920 ± 4.230 | 83.849 ± 4.218 | 84.684 ± 4.137 |
| Education (years) | 11.192 ± 3.035 | 12.452 ± 4.024 | -4.460[a] | <0.001* | 11.117 ± 2.871 | 12.732 ± 4.125 | 11.392 ± 3.156 | 12.985 ± 4.081 |
| BMI | 26.242 ± 4.222 | 27.205 ± 3.903 | -2.862[a] | 0.004* | 26.562 ± 4.336 | 27.173 ± 3.790 | 26.517 ± 4.191 | 26.769 ± 3.391 |
| HWR | 0.852 ± 0.070 | 0.954 ± 0.055 | -18.808[a] | <0.001* | 0.854 ± 0.057 | 0.953 ± 0.050 | 0.858 ± 0.059 | 0.940 ± 0.058 |
| TC (mmol/L) | 5.047 ± 0.987 | 4.589 ± 0.937 | 5.223 | <0.001* | 5.044 ± 0.922 | 4.580 ± 0.960 | 4.964 ± 0.800 | 4.559 ± 0.881 |
| Triglycerides (mmol/L) | 1.057 ± 0.492 | 1.095 ± 0.567 | -0.675[a] | 0.499 | 1.042 ± 0.492 | 1.078 ± 0.537 | 1.018 ± 0.467 | 1.071 ± 0.564 |
| HDL (mmol/L) | 1.586 ± 0.416 | 1.303 ± 0.410 | 7.554[a] | <0.001* | 1.574 ± 0.414 | 1.295 ± 0.412 | 1.570 ± 0.372 | 1.285 ± 0.355 |
| LDL (mmol/L) | 2.978 ± 0.902 | 2.789 ± 0.828 | 2.264[a] | 0.024* | 2.995 ± 0.862 | 2.809 ± 0.862 | 2.930 ± 0.775 | 2.787 ± 0.817 |
| SBP | 143.157 ± 19.558 | 148.699 ± 18.764 | -2.944[a] | 0.003* | 140.954 ± 19.898 | 144.087 ± 19.418 | 139.256 ± 20.372 | 140.009 ± 20.579 |
| DBP | 82.664 ± 9.420 | 84.133 ± 10.616 | -1.444[a] | 0.149 | 81.638 ± 11.054 | 81.556 ± 10.457 | 80.119 ± 11.807 | 78.953 ± 13.208 |
| Pulse rate | 72.373 ± 10.962 | 68.248 ± 11.928 | 4.218[a] | <0.001* | 71.480 ± 11.353 | 67.535 ± 11.429 | 77.435 ± 86.740 | 65.000 ± 11.322 |
| CVD score | 3.952 ± 3.478 | 4.191 ± 2.473 | -0.874[a] | 0.383 | 3.701 ± 3.560 | 3.686 ± 2.794 | 3.162 ± 3.513 | 2.850 ± 2.739 |
| HT (Yes:No) | 170:115 | 130:103 | 0.191[b] | 0.662 | 130:79 | 110:83 | 83:48 | 70:54 |
| Diabetes (Yes:No) | 24:261 | 30:203 | 6.429[b] | 0.011* | 17:192 | 35:158 | 8:123 | 19:105 |
| Ever smoking regularly (Yes:No) | 122:163 | 149:87 | 21.206[b] | <0.001* | -[c] | -[c] | 5:129 | 3:121 |
| Alcohol consumption (High:Low) | 130:156 | 162:74 | 29.922[b] | <0.001* | 98:111 | 127:66 | 59:75 | 82:43 |
| TWMH (mm$^3$) | 18358.159 ± 23508.124 | 14464.212 ± 13143.762 | 2.340[a] | 0.020* | 21777.476 ± 30065.807 | 17781.295 ± 15419.174 | 25117.569 ± 29924.230 | 23988.021 ± 21262.817 |
| PVWMH (mm$^3$) | 10479.507 ± 9623.665 | 8990.999 ± 6991.410 | 1.593[a] | 0.112 | 11410.184 ± 9423.738 | 10434.109 ± 7340.914 | 13023.159 ± 9329.597 | 12941.367 ± 9713.258 |
| DWMH (mm$^3$) | 7789.037 ± 15841.933 | 5401.402 ± 7656.987 | 2.748[a] | 0.006* | 10280.075 ± 22713.53 | 7254.555 ± 9752.148 | 11427.076 ± 19992.770 | 10621.818 ± 15462.544 |
| Processing speed | -0.0047±1.0379 | -0.061±1.024 | 0.675[a] | 0.500 | -0.058±1.105 | -0.173±1.332 | -0.383±1.274 | -0.606±1.474 |
| Language | -0.135±1.095 | -0.051±1.141 | -0.925[a] | 0.355 | -0.258±1.063 | -0.170±1.092 | -0.425±1.285 | -0.3699±1.184 |
| Executive function | -0.049±1.058 | 0.015±1.021 | -0.746[a] | 0.455 | -0.130±1.219 | -0.170±1.206 | -0.347±1.330 | -0.470±1.629 |
| **Visuospatial** | -0.152±0.939 | 0.200±1.039 | -4.363[a] | <0.001* | -0.086±1.066 | 0.180±1.077 | -0.299±1.065 | -0.086±1.166 |
| **Memory** | 0.152±1.052 | -0.293±0.987 | 5.324[a] | <0.001* | 0.133±1.107 | -0.425±1.033 | 0.0167±1.256 | -0.566±1.158 |
| **Global cognition** | -0.069±1.069 | -0.063±1.076 | -0.069[a] | 0.945 | -0.130±1.138 | -0.203±1.190 | -0.401±1.271 | -0.632±1.518 |

N = number of participants; TWMH = Total White Matter Hyperintensities; PVWMH = Periventricular White Matter Hyperintensities; DWMH = Deep White Matter Hyperintensities; CVD = cardiovascular disease; F = Female; M = Male; BMI = Body Mass Index; HDL = high density lipoprotein; HWR = Hip to waist ratio; HT = Hypertension; SBP = systolic blood pressure; DBP = diastolic blood pressure; LDL = lower density lipoprotein; Std. = Standard deviation; TC = total cholesterol. [a] Independent-sample *t*-test to detect differences in continuous risk factors between men and women;. [b] Chi-square test to detect differences in binary risk factors between men and women; [c] Data not available. * statistically significant at alpha = 0.05.

Table 2: The effects of vascular risk factors on WMH progression.

| Risk factors | variable in LMM | TWMH | | PVWMH | | DWMH | |
|---|---|---|---|---|---|---|---|
| | | Exp(B) | p-value | Exp(B) | p-value | Exp(B) | p-value |
| BMI | TIS×BMI | 0.999 | 0.383 | 0.999 | 0.244 | 0.998 | 0.216 |
| HWR | TIS×HWR | 1.065 | 0.367 | 0.895 | 0.037 | 1.223 | 0.047 |
| TC (mmol/L) | TIS×TC | 1.000 | 0.864 | 1.005 | 0.144 | 0.996 | 0.653 |
| Triglycerides (mmol/L) | TIS× triglycerides | 1.000 | 0.906 | 0.988 | 0.076 | 1.010 | 0.479 |
| HDL (mmol/L) | TIS×HDL | 1.002 | 0.799 | 1.018 | 0.045 | 0.988 | 0.571 |
| LDL (mmol/L) | TIS×LDL | 0.999 | 0.980 | 1.004 | 0.276 | 0.995 | 0.657 |
| SBP (10 mmHg) | TIS×Systolic BP | 0.995 | 0.058 | 0.995 | 0.022 | 0.990 | 0.016 |
| DBP (10 mmHg) | TIS×Diastolic BP | 0.997 | 0.525 | 0.999 | 0.687 | 0.983 | 0.031 |
| Pulse rate | TIS×Pulse rate | 1.000 | 0.741 | 1.000 | 0.598 | 0.999 | 0.824 |
| HT | TIS×HT | 0.980 | 0.069 | 0.988 | 0.145 | 0.969 | 0.055 |
| Diabetes | TIS×Diabetes | 0.996 | 0.792 | 0.999 | 0.954 | 0.999 | 0.980 |
| Smoking | TIS×smoking | 1.004 | 0.669 | 1.008 | 0.349 | 0.999 | 0.974 |
| Alcohol | TIS×alcohol | 1.008 | 0.454 | 1.010 | 0.238 | 0.997 | 0.858 |
| CVD risk scores | TIS×CVD risk score | 0.998 | 0.229 | 0.998 | 0.071 | 0.997 | 0.143 |

LMM = Linear Mixed Model; BMI = body mass index; SBP = systolic blood pressure; DBP = diastolic blood pressure; CVD = cardiovascular diseases; HT = hypertension; HWR = hip-to-waist ratio; TIS = time-in-study; TWMH = total white matter hyperintensities; DWMH = deep white matter hyperintensities; PVWMH = periventricular white matter hyperintensities. None of the associations between vascular risk factors and WMH progression survived Bonferroni correction. Exp(B) is the exponentiated unstandardised coefficient B from the linear mixed models.

Table 3: Sex moderation effects on the associations between vascular risk factors and changes in WMH volumes.

| Risk factors | Variable in LMM | TWMH | | PVWMH | | DWMH | |
|---|---|---|---|---|---|---|---|
| | | Exp(B) | p-value | Exp(B) | p-value | Exp(B) | p-value |
| BMI | TIS×BMI×sex | 1.001 | 0.623 | 1.001 | 0.690 | 1.001 | 0.742 |
| HWR | TIS×HWR×sex | 1.087 | 0.675 | 1.189 | 0.264 | 1.027 | 0.923 |
| TC | TIS×TC×sex | 1.004 | 0.659 | 1.002 | 0.731 | 1.005 | 0.786 |
| Triglycerides | TIS×Triglycerides×sex | 1.005 | 0.762 | 1.011 | 0.374 | 0.987 | 0.650 |
| HDL | TIS×HDL×sex | 1.003 | 0.899 | 1.008 | 0.666 | 1.024 | 0.568 |
| LDL | TIS×LDL×sex | 1.004 | 0.690 | 1.001 | 0.838 | 1.002 | 0.894 |
| SBP | TIS×SBP×sex | 0.989 | 0.040 | 0.995 | 0.233 | 0.982 | 0.020 |
| | TIS×SBP in women [a] | 1 | 0.896 | 0.998 | 0.401 | 0.998 | 0.661 |
| | TIS×SBP in men [a] | 0.988 | 0.004 | 0.993 | 0.016 | 0.979 | <0.001 |
| DBP | TIS×DBP×sex | 0.982 | 0.088 | 0.993 | 0.369 | 0.970 | 0.054 |
| Pulse rate | TIS×Pulse rate×sex | 1.001 | 0.449 | 1.000 | 0.637 | 1.001 | 0.362 |
| HT | TIS×HT×sex | 0.999 | 0.956 | 1.002 | 0.897 | 0.982 | 0.570 |
| Diabetes | TIS×Diabetes×sex | 0.984 | 0.669 | 0.979 | 0.519 | 0.955 | 0.392 |
| Smoking | TIS×Smoking×sex | 0.995 | 0.802 | 0.985 | 0.400 | 1.016 | 0.634 |
| Alcohol | TIS×Alcohol×sex | 0.980 | 0.403 | 0.994 | 0.737 | 0.984 | 0.641 |
| CVD risk score | TIS×CVD risk score×sex | 1.122 | 0.115 | 0.997 | 0.309 | 0.985 | 0.006* |
| | TIS×CVD risk score in women [a] | 1.000 | 0.944 | 0.999 | 0.438 | 1.001 | 0.704 |
| | TIS×CVD risk score in men [a] | 0.994 | 0.044 | 0.996 | 0.067 | 0.987 | 0.002 |

[a] follow-up analysis conducted when the TIS×risk factor×sex interaction term is significant to identify the effects in men and women separately. Exp(B) is the exponentiated coefficient was calculated from unstandardised B generated from the linear mixed models with transformed WMH. BMI = body mass index; SBP = systolic blood pressure; DBP = diastolic blood pressure; CVD = cardiovascular diseases; HT = hypertension; HWR = hip-to-waist ratio; TIS = time-in-study; LMM = Linear Mixed Model; TWMH = total white matter hyperintensities; DWMH = deep white matter hyperintensities; PVWMH = periventricular white matter hyperintensities. * survive Bonferroni correction alpha = 0.00833.

Table 4: The associations between WMH progression and changes in cognitive performance (both between- and within-subject effects)

| Cognition | Between/Within-subject effect | TWMH [a] | | PVWMH [a] | | DWMH [a] | |
|---|---|---|---|---|---|---|---|
| | | Estimate | P-value | Estimate | P-value | Estimate | P-value |
| Processing speed | between | -0.006 | 0.002* | -0.014 | 0.005 | -0.008 | 0.004 |
| | within | -0.004 | 0.183 | -0.018 | 0.073 | -0.003 | 0.376 |
| Language | between | -0.002 | 0.318 | -0.007 | 0.157 | -0.001 | 0.557 |
| | within | -0.002 | 0.368 | -0.012 | 0.123 | -0.001 | 0.626 |
| Executive function | between | -0.007 | 0.001* | -0.016 | 0.001* | -0.009 | 0.004 |
| | within | -0.005 | 0.143 | -0.022 | 0.040 | -0.004 | 0.266 |
| Visuospatial | between | -0.002 | 0.217 | -0.007 | 0.111 | -0.002 | 0.412 |
| | within | -0.002 | 0.447 | -0.029 | 0.001* | 0.000 | 0.919 |
| Memory | between | -0.002 | 0.379 | -0.006 | 0.156 | -0.001 | 0.682 |
| | within | -0.005 | 0.029 | -0.027 | 0.001* | -0.004 | 0.180 |
| Global cognition | between | -0.005 | 0.007 | -0.014 | 0.003 | -0.006 | 0.033 |
| | within | -0.004 | 0.056 | -0.026 | 0.001* | -0.003 | 0.244 |

TWMH = total white matter hyperintensities; DWMH = deep white matter hyperintensities; PVWMH = periventricular white matter hyperintensities; mL = millilitre. * survive Bonferroni correction at alpha = 0.00277. [a] WMH volumes were converted from $mm^3$ to mL for better illustrating the effect sizes.

Table 5: Sex moderation effects on the associations between WMH volumes (between-/within-subject effects) and cognitive performance

| Cognition | Effect | | TWMH [a] | | PVWMH [a] | | DWMH [a] | |
|---|---|---|---|---|---|---|---|---|
| | | | Estimate | P-value | Estimate | P-value | Estimate | P-value |
| Processing speed | Between-subject | sex×WMH interaction | 0.002 | 0.756 | 0.009 | 0.348 | 3.000E-4 | 0.966 |
| | Within-subject | sex×WMH interaction | -0.007 | 0.176 | -0.019 | 0.231 | -0.007 | 0.293 |
| Language | Between-subject | sex×WMH interaction | -0.004 | 0.319 | -0.004 | 0.696 | -0.008 | 0.269 |
| | Within-subject | sex×WMH interaction | -0.001 | 0.766 | 0.009 | 0.476 | -0.002 | 0.604 |
| Executive function | Between-subject | sex×WMH interaction | 0.008 | 0.101 | 0.032 | 0.002* | 0.004 | 0.609 |
| | | Effect for women | - | - | -0.026 | <0.001 | - | - |
| | | Effect for men | - | - | 0.006 | 0.494 | - | - |
| | Within-subject | sex×WMH interaction | -0.003 | 0.537 | -0.025 | 0.156 | -0.001 | 0.845 |
| Visuospatial | Between-subject | sex×WMH interaction | 0.001 | 0.851 | 0.001 | 0.839 | 0.001 | 0.800 |
| | Within-subject | sex×WMH interaction | -0.011 | 0.022 | -0.042 | 0.002* | -0.008 | 0.176 |
| | | Effect for women | 0.002 | 0.493 | -0.007 | 0.480 | - | - |
| | | Effect for men | -0.008 | 0.027 | -0.050 | <0.001 | - | - |
| Memory | Between-subject | sex×WMH interaction | 0.002 | 0.499 | 0.006 | 0.510 | 0.005 | 0.444 |
| | Within-subject | sex×WMH interaction | -0.003 | 0.438 | -0.014 | 0.273 | -0.002 | 0.616 |
| Global cognition | Between-subject | sex×WMH interaction | 0.002 | 0.634 | 0.012 | 0.214 | -1.012E-5 | 0.999 |
| | Within-subject | sex×WMH interaction | -0.008 | 0.052 | -0.022 | 0.078 | -0.008 | 0.137 |

TWMH = total white matter hyperintensities; DWMH = deep white matter hyperintensities; PVWMH = periventricular white matter hyperintensities.

* Indicates those survive Bonferroni correction alpha = 0.00277. [a] WMH volumes were converted from $mm^3$ to mL to better illustrate the effect sizes.

# Figures

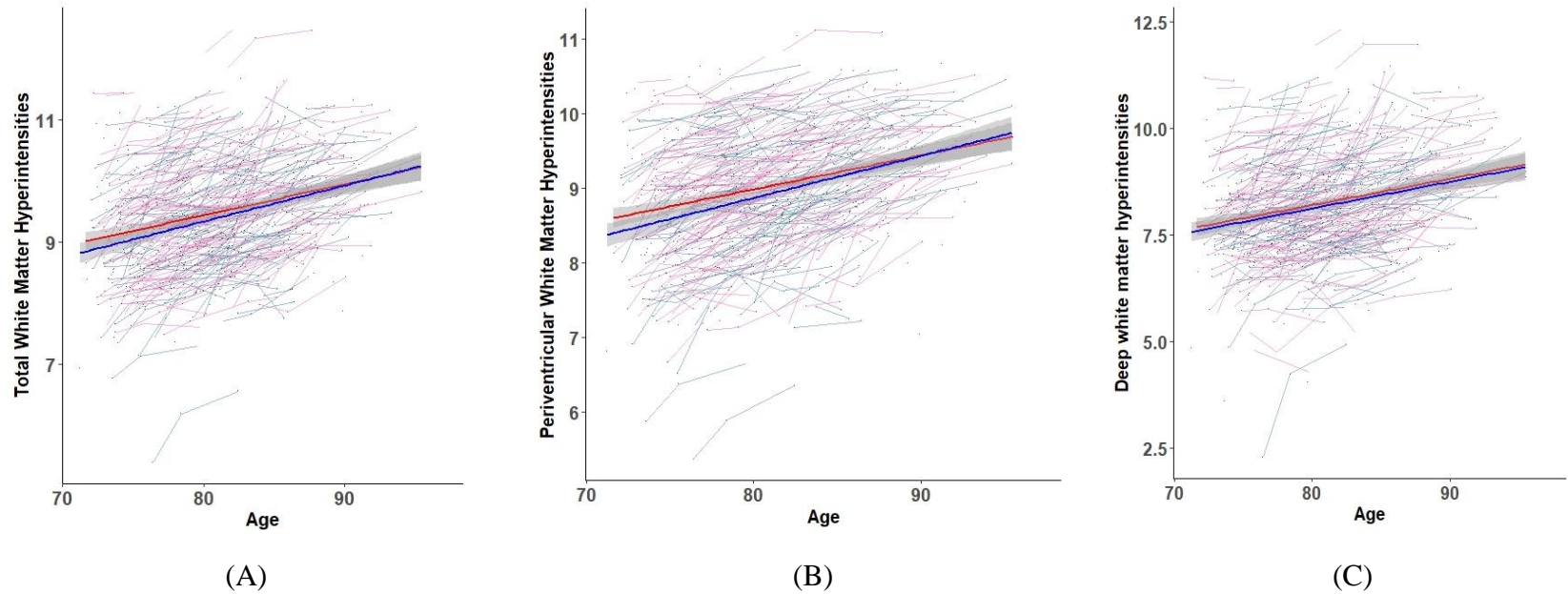

(A)　　　　　　　　　　　　　　　　(B)　　　　　　　　　　　　　　　　(C)

*Figure 1: Longitudinal changes of total (A), periventricular (B) and deep (C) white matter hyperintensities volumes (natural logarithm) over 6 years in men and women.* Individual trajectories of WMH volume changes over 6 years were shown by blue lines for men, or pink lines for women. The bold blue and red lines were average WMH of all time-points in men and women, respectively.

Supplementary Table 1: Differences in risk factors and cognition between men and women at 2- and 6-year follow-up.

| Parameter | 2-year follow-up | | | | 6-year follow-up | | | |
|---|---|---|---|---|---|---|---|---|
| | Female | Male | Statistics | p-value | Female | Male | Statistics | p-value |
| N | 211 | 193 | - | - | 135 | 125 | - | - |
| Age (years) | 80.505 ± 4.680 | 80.920 ± 4.230 | -0.215[a] | 0.830 | 83.849 ± 4.218 | 84.684 ± 4.137 | 0.776[a] | 0.438 |
| Education (years) | 11.117 ± 2.871 | 12.732 ± 4.125 | -5.176[a] | <0.001* | 11.392 ± 3.156 | 12.985 ± 4.081 | -4.762[a] | <0.001* |
| BMI | 26.562 ± 4.336 | 27.173 ± 3.790 | -2.716[a] | 0.007* | 26.517 ± 4.191 | 26.769 ± 3.391 | -2.014[a] | 0.044* |
| HWR | 0.854 ± 0.057 | 0.953 ± 0.050 | -23.025[a] | <0.001* | 0.858 ± 0.059 | 0.940 ± 0.058 | -15.139[a] | <0.001* |
| TC (mmol/L) | 5.044 ± 0.922 | 4.580 ± 0.960 | -[c] | -[c] | 4.964 ± 0.800 | 4.559 ± 0.881 | -[c] | -[c] |
| Triglycerides (mmol/L) | 1.042 ± 0.492 | 1.078 ± 0.537 | -[c] | -[c] | 1.018 ± 0.467 | 1.071 ± 0.564 | -[c] | -[c] |
| HDL (mmol/L) | 1.574 ± 0.414 | 1.295 ± 0.412 | -[c] | -[c] | 1.570 ± 0.372 | 1.285 ± 0.355 | -[c] | -[c] |
| LDL (mmol/L) | 2.995 ± 0.862 | 2.809 ± 0.862 | -[c] | -[c] | 2.930 ± 0.775 | 2.787 ± 0.817 | -[c] | -[c] |
| SBP | 140.954 ± 19.898 | 144.087 ± 19.418 | -2.873[a] | 0.004* | 139.256 ± 20.372 | 140.009 ± 20.579 | -0.173[a] | .863 |
| DBP | 81.638 ± 11.054 | 81.556 ± 10.457 | 0.042[a] | 0.967 | 80.119 ± 11.807 | 78.953 ± 13.208 | 1.760[a] | 0.079 |
| Pulse rate | 71.480 ± 11.353 | 67.535 ± 11.429 | 4.517[a] | <0.001* | 77.435 ± 86.740 | 65.000 ± 11.322 | 3.373[a] | 0.001* |
| CVD score | 3.701 ± 3.560 | 3.686 ± 2.794 | -1.328[a] | 0.185 | 3.162 ± 3.513 | 2.850 ± 2.739 | 0.883[a] | 0.378 |
| HT (Yes:No) | 130:79 | 110:83 | 0.272[b] | 0.602 | 83:48 | 70:54 | 1.438[b] | 0.230 |
| Diabetes (Yes:No) | 17:192 | 35:158 | 21.300[b] | <0.001* | 8:123 | 19:105 | 16.459[b] | <0.001* |
| Ever smoking regularly (Yes:No) | -[c] | -[c] | - | - | 5:129 | 3:121 | 0.685[b] | 0.408 |
| Alcohol consumption (High:Low) | 98:111 | 127:66 | 35.507[b] | <0.001* | 59:75 | 82:43 | 22.047[b] | 0.001* |
| TWMH (mm³) | 21777.476 ± 30065.807 | 17781.295 ± 15419.174 | 1.582[a] | 0.115 | 25117.569 ± 29924.230 | 23988.021 ± 21262.817 | 0.416[a] | 0.678 |
| PVWMH (mm³) | 11410.184 ± 9423.738 | 10434.109 ± 7340.914 | 0.934[a] | 0.351 | 13023.159 ± 9329.597 | 12941.367 ± 9713.258 | 0.379[a] | 0.705 |
| **DWMH (mm³)** | 10280.075 ± 22713.53 | 7254.555 ± 9752.148 | 1.713[a] | 0.088 | 11427.076 ± 19992.770 | 10621.818 ± 15462.544 | 0.315[a] | 0.753 |
| **Processing speed** | -0.058±1.105 | -0.173±1.332 | 1.075[a] | 0.282 | -0.383±1.274 | -0.606±1.474 | 1.614[a] | 0.107 |
| **Language** | -0.258±1.063 | -0.170±1.092 | -0.942[a] | .347 | -0.425±1.285 | -0.3699±1.184 | -0.464[a] | 0.643 |
| **Executive function** | -0.130±1.219 | -0.170±1.206 | 0.376[a] | 0.707 | -0.347±1.330 | -0.470±1.629 | 0.814[a] | 0.416 |
| **Visuospatial** | -0.086±1.066 | 0.180±1.077 | -2.895[a] | 0.004 | -0.299±1.065 | -0.086±1.166 | -1.952[a] | 0.052 |
| **Memory** | 0.133±1.107 | -0.425±1.033 | 5.999 | <0.001* | 0.0167±1.256 | -0.566±1.158 | 4.852 | <0.001* |
| **Global cognition** | -0.130±1.138 | -0.203±1.190 | 0.725 | 0.469 | -0.401±1.271 | -0.632±1.518 | 1.700 | 0.090 |

N = number of participants; TWMH = Total White Matter Hyperintensities; PVWMH = Periventricular White Matter Hyperintensities; DWMH = Deep White Matter Hyperintensities; CVD = cardiovascular disease; F = Female; M = Male; BMI = Body Mass Index; HDL = high density lipoprotein; HWR = Hip to waist ratio; HT = Hypertension; SBP = systolic blood pressure; DBP = diastolic blood pressure; LDL = lower density lipoprotein; Std. = Standard deviation; TC = total cholesterol. [a] Independent-sample *t*-test to detect differences in continuous risk factors between men and women;. [b] Chi-square test to detect differences in binary risk factors between men and women; [c] Data not available. * statistically significant at alpha = 0.05.

Supplementary Table 2: Sex moderation effects on the association between CVD risk scores and changes in WMH volumes in younger (⩽ 79 years) and older (> 79 years) participants.

| Risk factors | Variable in LMM | DWMH Exp(B) | DWMH p-value |
|---|---|---|---|
| **CVD risk score (younger participants)** | TIS×CVD risk score×sex | 1.009 | 0.203 |
| | TIS×CVD risk score in women [a] | 1.001 | 0.781 |
| | TIS×CVD risk score in men [a] | 0.992 | 0.173 |
| **CVD risk score (older participants)** | TIS×CVD risk score×sex | 1.020 | 0.016 |
| | TIS×CVD risk score in women [a] | 1.001 | 0.912 |
| | TIS×CVD risk score in men [a] | 0.981 | 0.003 |

[a] follow-up analysis conducted when the TIS×risk factor×sex interaction term is significant to identify the effects in men and women separately. Exp(B) is the exponentiated coefficient was calculated from unstandardised B generated from the linear mixed models. CVD = cardiovascular diseases; TIS = time-in-study; LMM = Linear Mixed Model; DWMH = deep white matter hyperintensities.